\documentclass[12pt,a4paper]{article}
\usepackage[parfill]{parskip}    
\usepackage[english]{babel}
\usepackage[applemac]{inputenc}
\usepackage{fullpage}
\usepackage{graphicx}
\usepackage{amssymb}
\usepackage{setspace}
\usepackage{rsc}

\begin{document}

DOI: 10.1002/smll.((please add manuscript number)) 

\textbf{Nanodiamond as a vector for siRNA delivery to Ewing sarcoma cells}

\emph{Anna Alhaddad, Marie-Pierre Adam, G\'eraldine Dantelle, Sandrine Perruchas, Thierry Gacoin, Christelle Mansuy, Solange Lavielle, Claude Malvy, Fran\c cois Treussart**, and Jean-Rémi Bertrand*}\\

[*] \hspace{10 mm} Dr.~J.-R.~Bertrand, A.~Alhaddad, Prof.~C.~Malvy \\
Laboratoire de Vectorologie et Th\'erapeutiques Anticanc\'ereuses, CNRS UMR 8203, Universit\'e Paris Sud 11, Institut de Canc\'erologie Gustave Roussy, Villejuif 94805 cedex, France\\
E-mail: jean-remi.bertrand@igr.fr\\

[**] \hspace{8 mm} Prof.~F.~Treussart, M.-P.~Adam, Dr. J.~Botsoa \\
Laboratoire de Photonique Quantique de Mol\'eculaire, CNRS UMR 8537, \'Ecole Normale Sup\'erieure de Cachan, F-94235 Cachan cedex, France \\
E-mail: francois.treussart@ens-cachan.fr\\

\hspace{15 mm} Dr.~G.~Dantelle, Dr.~S.~Perruchas, Dr.~T.~Gacoin\\
 Laboratoire de Physique de la Mati\`ere Condens\'ee, CNRS UMR 7643, \'Ecole Polytechnique, F-91120 Palaiseau, France\\
 
\hspace{15 mm} Dr.~C.~Mansuy, Prof.~S.~Lavielle\\
Laboratoire des BioMol\'ecules\\
CNRS UMR 7203\\ ENS, D\'epartement de Chimie, \'Ecole Normale Sup\'erieure, 24 rue Lhomond, F-75231 Paris cedex, France and \\ UPMC Paris~6, Universit\'e Pierre et Marie Curie, 4, Place Jussieu, F-75005 Paris, France\\

Supporting Information: 
\begin{itemize}
\item Viabillity of cells expressing EWS-Fli1 oncogene after incubation with siRNA:ND
\item Determination of theoptimal mass of ND-polycation for a sufficient amount of siRNA:ND-PAH/PEI internalized in NIH/3T3 EF cells in culture
\end{itemize}
is available on the WWW under http://www.small-journal.com or from the author.


\doublespacing

\newpage

\section*{Abstract}

 We investigated the ability of diamond nanoparticles (nanodiamonds, NDs) to deliver small interfering RNA (siRNA)  in Ewing sarcoma cells, in the perspective of {\it in vivo} anti-cancer nucleic acid drug delivery. siRNA was adsorbed onto NDs previously coated with cationic polymer. Cell uptake of NDs has been demonstrated by taking advantage of NDs intrinsic fluorescence coming from embedded color center defects. Cell toxicity of these coated NDs was shown to be low. Consistent with the internalization efficacy, we have shown a specific inhibition of EWS/Fli-1 gene expression at the mRNA and protein level by the ND vectorized siRNA in a serum containing medium.

\section{Introduction}
In the anti-cancer drug delivery domain, nanotechnologies are a promising tool for future applications such as a more favourable tissue distribution and a decrease of the toxicity~\cite{Qi:2010bj,*Caliceti:2003dj,*Pouton:1998vd,*Nakada:1996ur}.Various types of nanoparticles have been developed~\cite{Tang:2010ue} such as liposomes, lipidic micelles, dendrimers and polymers. Active molecules are either covalently bound to the nanoparticles~\cite{Huang:2007ej,Liu:2009gp}, adsorbed onto their surface~\cite{Zhang:2009hc,Kurosaki:2009gq} or encapsulated~\cite{DiazLopez:2010ig}. 

The requirements for the efficacy of a vector are (i) a large quantity of active molecules per vector entity, (ii) the capacity to reach the targeted cells, (iii) the capacity to deliver the active molecule to the target in these cells. As far as side effects are concerned the nanoparticle needs to have a low toxicity, to be bio-compatible, and to be quickly eliminated from the body after their action has taken place. Therefore, in order to find the most potent compounds a large variety of vectors needs to be investigated. For instance the use of mineral nano-object as carrier for drug delivery has been proposed including silica, metals, and carbon~\cite{Horcajada:2010ce,Huang:2010cm,Simovic:2010di,Liu:2009gp}. 

In this work we used diamond nanocrystals (nanodiamonds)~\cite{Barnard:2009eo,Schrand:2009wn} as a cell targeting bi-functional device, serving as a drug delivery vehicle and a fluorescent label at the same time. Nanodiamond display attractive properties such as a small size, down to less than 10~nm, various possibilities of surface funtionnalization~\cite{VaijayanthimalaV:2009ca} and a perfectly stable photoluminescence well adapted for studying intracellular traffic and localization. The fluorescence properties results from the creation of nitrogen-vacancy (NV) color centers inside the nanodiamond matrix. The nitrogen is naturally present in synthetic diamond produced under high-pressure and high temperature conditions. The vacancy is created by high-energy ion beam irradiation, and the NV complex is formed after thermal annealing. Various approaches have been recently reported for the production of bright fluorescent nanodiamonds (fNDs) in quantities sufficiently large for the need of bio-applications~\cite{Boudou:2009jl,Chang:2008ia,Dantelle:2010hc}. 
Compared to semiconductor nanocrystals (QDots), NV color centers contained into NDs present a perfectly stable emission (no blinking, no bleaching), with a similar brightness at saturation of their fluorescence~\cite{Faklaris:2009ec}. They can therefore be used for long-term tracking in cells and organisms~\cite{Mohan:2010bf,Chang:2008ia}. 

Furthermore, NDs seem to be better tolerated by cells than other carbon nanomaterials~\cite{Schrand:2007eu}.  Fluorescent NDs  introduced into {\it Caenorhabditis elegans}  worm, by either feeding or microinjection into the gonads proved to be stable  nontoxic and did not cause any detectable stress to the worms~\cite{Mohan:2010bf}. 
In mice, intravenous injection of 50~nm size NDs led to longterm (28~days) entrapment in the liver and the lung~\cite{YUAN:2009dw}, but no mice showed any symptoms of abnormality. However, further toxicological studies are necessary due to the accumulation. Moreover, when instilled in the mouse trachea 4~nm NDs did not show evident adverse effects in the lungs, owing to their efficient elimination by macrophages~\cite{Yuan:2009bd}.  
\enlargethispage{-65.1pt}
Considering these encouraging results in cellular and small animal models, \emph{in vitro} experiments have been recently performed with NDs to show their capacity to deliver anticancer drugs, in cell~\cite{Huang:2007ej,Guan:2010iw} and in mouse model of liver and mammary cancers where ND delivery enhances the chemotherapeutic efficiency of the drug~\cite{Chow:2011hf}.

Recent studies have shown the potential of nanodiamonds coated by cationic polymers to deliver genes into the cells~\cite{Zhang:2009hc}, including siRNA~\cite{Chen:2010kq}, without adverse effects, and with higher efficiency than the standard lipofectamine transfecting agent in physiological conditions.
siRNA are short double strand RNAs inducing gene-silencing activity by interfering with mRNA in the cell cytoplasm by triggering a sequence specific cleavage at the level of sequence recognition on target mRNA after association to the RISC complex~\cite{Fattal:2009dm}.

In this work we establish the relevance of such a strategy in a therapeutic context. We used NDs as a cargo to deliver siRNA in human cell lines originating from Ewing Sarcoma tumors. 
Ewing Sarcoma, a genetic disease considered as the most frequent bone cancer in children, is induced in 90\% of case by chromosomal translocation at the level of chromosomes t(11,22). This results in EWS-Fli1 fusion gene, which is  finally expressed as a chimeric protein~\cite{Zoubek:1996tj}. Treatment of this cancer involves surgical resection, radiotherapy and chemotherapy. New strategies based on the inhibition of EWS-Fli1 gene by antisense oligonucleotides or siRNA targeting the EWS-Fli1 junction at the mRNA level have been  developed~\cite{Ramon:2008wy}. siRNA were shown to trigger the EWS-Fli1 mRNA cleavage and showed an inhibition in vitro of EWS-Fli1 expression and in vivo of tumor growth~\cite{Maksimenko:2003vz,Maksimenko:2003dn,Tanaka:1997fl,Toub:2006bd}. 
Despite their high \emph{in vitro} efficiency siRNAs are quickly degraded in physiological fluids and have a poor capacity as polyanions to enter cells. Their \emph{in vivo} efficiency can therefore be greatly improved by the use of cargo delivery systems~\cite{Elhamess:2009er,deMartimprey:2008kq,Toub:2006bd}.

We developed cationic NDs able to bind siRNAs at their surface. We compared two types of polymers capable of interacting with anionic oligonucleotides polyethylene-imine (PEI) and polyallylamine hydrochloride (PAH).
PEI  is currently used as an oligonucleotide transfection agent. However, the water-soluble PEI does not form siRNA polyplexes stable enough in extracellular media for effective delivery, but the system consisting in noncovalent coating of PEI to the surface of a nanoparticle has proved to be a very efficient DNA and siRNA delivery device~\cite{Xia:2009kb,Zhang:2009hc,Chen:2010kq}. 
We also used PAH coating because it was shown to form a reproducible complex leading to a stable aqueous suspension with a low cell toxicity~\cite{Vial:2008jw}.

Herein, we  analyzed  the data obtained with these two types of coating, using fluorescent NDs, in terms of their capacity (i) to bind to siRNA, (ii) to allow its cell uptake, and (iii) to inhibit EWS-Fli1 expression in cell culture at both the mRNA and protein levels.

\section{Results and discussion}

\subsection{Characterization of cationic nanodiamonds}

Both types of ND (as received powder, or electron irradiated and annealed ones) have undergone an oxidative treatment to remove the graphitic shell, resulting in a surface rich in carboxylic groups, and forming anionic particles. Cationic NDs were produced by surface coating with amino rich polymers: Poly(allylamine hydrochlolide)(PAH) red, following our previous work procedure~\cite{Vial:2008jw}, or Polyethileneimine (PEI). As indicated in {\bf Table~\ref{table:01}}, the pH are 5.6 and 6.7 for ND-PAH and ND-PEI aqueous suspensions respectively. ND zeta potential shifts from  -27~mV before polymer coating  to +31~mV or +26~mV after coating by PAH or PEI, respectively. This surface coating is associated to an increase of their size from 50~nm to 130~nm and 120 nm respectively, which is most likely due to the formation of small aggregates during the coating. When the particle charge turns from negative to positive value, repulsion between particles is lower, and aggregation is generally observed. It is well known that nanoparticles have a propensity to form strongly bound aggregates hard to break even by strong sonication. However beads assisted sonication has recently been successful for de-agglomeration and simultaneous covalent functionalization of nanodiamonds at their primary particle level~\cite{Liang:2009jj}.

FT-IR spectroscopy reported in a previous study~\cite{Vial:2008jw} confirms the adsorption of PAH on nanodiamonds. For PAH and PEI coated NDs, the quantity of the amino group was determined by the Kaiser methods to be 274 and 173~$\mu$mole per gram of NDs, respectively. However, in the latter case this amount is underestimated, as the tertiary amines present on a branched PEI (25\% of the total amino groups) do not react with the ninhydrin.
 	
\subsection{Adsorption of siRNA onto cationic ND: determination of the concentrations at saturation}

\begin{figure}[ht!]
\begin{center}
\includegraphics[width=0.8\textwidth]{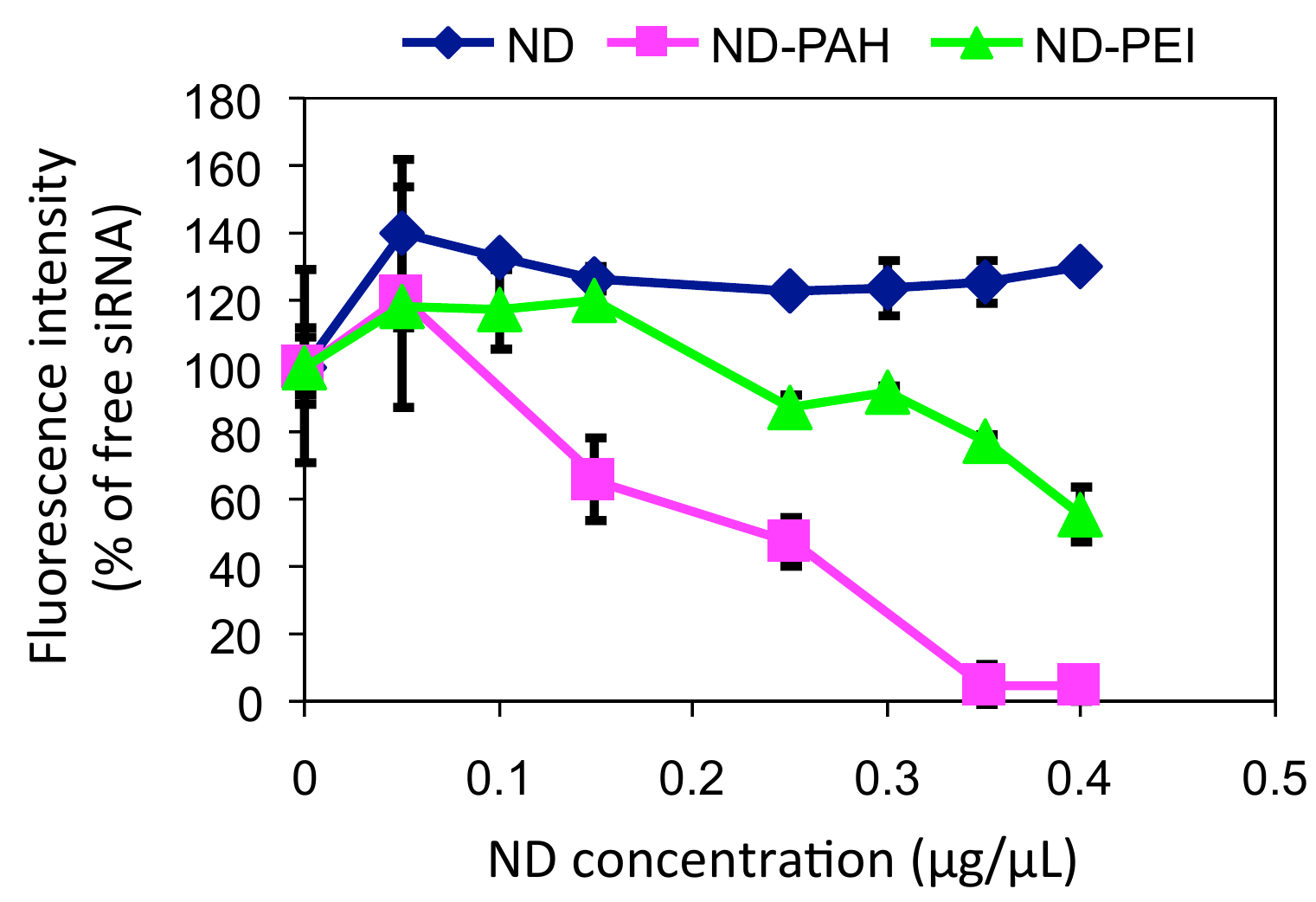}
\end{center}
\caption{\doublespacing Titration of the binding of siRNA to NDs surface. Free siRNA concentration is determined by ethidium bromide coloration after separation from NDs by centrifugation as described in the Experimental Section. Initial concentration of siRNA: 5 ng/$\mu$l, corresponding to 384~nM.}
\label{fig1}
\end{figure}

The siRNA adsorption capacity of the polymer-coated NDs has been measured by adding a siRNA solution at the smallest concentration compatible with the quantification method, i.e. 5 ng$\,\mu$l$^{-1}$ (corresponding to 384~nM) with increasing concentrations of NDs (0-0.4~$\mu$g$\,\mu$l$^{-1}$). The adsorption of siRNA do not lead to increase of hydrodynamic radius of the nanodiamond complex, despite a decrease of the zeta potential to about +10~mV. 
After centrifugation, the free nucleic acid concentration is determined by fluorescence of the supernatant fractions after ethidium bromide coloration. {\bf Figure~\ref{fig1}} shows the saturation curve obtained with as-received anionic NDs, and cationic PAH- or PEI-coated NDs.  A decrease of the free nucleic acid content is observed when cationic polymer-coated NDs are added to the siRNA with increasing concentrations, while no change happens with the as-received anionic NDs, as expected from charge repulsion. The saturation of siRNA adsorption onto ND-PAH happens at ND-PAH concentration of 0.35~$\mu$g$\,\mu$l$^{-1}$, corresponding to a siRNA:ND-PAH mass ratio of 1:70 and a phosphate to amino groups charge ratio of 1:6. 
In the case of ND-PEI solutions we also observe a decrease of the free siRNA upon increase of the ND concentration, but the siRNA charges are never fully compensated by ND-PEI positive charges in the concentration range explored. At the maximal ND concentration of 0.4~$\mu$g$\,\mu$l$^{-1}$, the mass ratio of free siRNA to ND is 1:120 associated to a 1:4 charge ratio. However, siRNA total adsorption onto ND-PEI can be extrapolated from a linear fit of the curve and corresponds to concentration of 0.7~$\mu$g$\,\mu$l$^{-1}$ associated to a saturation mass ratio of 1:140 and a phosphate to amino charge ratio of 1:8.
The conclusion is that at a given ND concentration, ND-PAH is able to adsorb a higher amount of siRNA than ND-PEI. Note that the slightly larger zeta potential of ND-PAH compared to that of ND-PEI also supports this conclusion.

\subsection{Cytotoxicity of the ND on NIH/3T3 cells}

\begin{figure}[ht!]
\begin{center}
\includegraphics[width=0.8\textwidth]{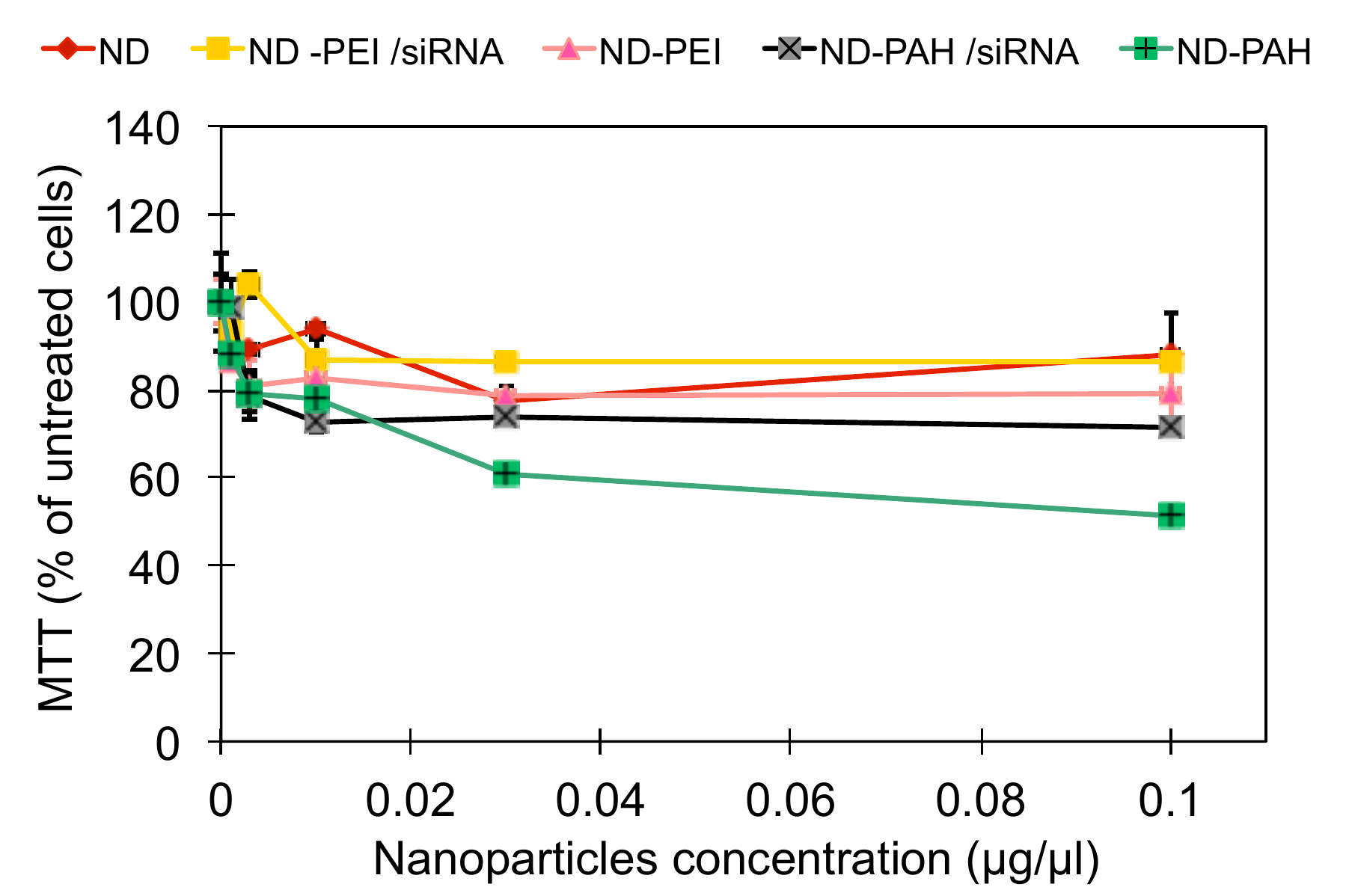}
\end{center}
\caption{\doublespacing Cell viability measurements by MTT assay using NIH/3T3. The cells were incubated with NDs and NDs coated with positive polymers with siRNA (50~nM) eventually bound to them for doses from $0.001-0.1~\mu$g$\,\mu$l$^{-1}$ for 24~h. Viability results are presented in percent as the ratio of MTT for treated cells on MTT for untreated cells.}
\label{fig2}
\end{figure}

In order to evaluate the toxicity of the nanodiamond, we have studied the cell viability of NIH/3T3 murine fibroblasts in the presence of NDs, NDs coated with the polycations (PAH or PEI), and ND-siRNA complexes. In the latter case, the typical concentration of 50~nM of siRNA was used~\cite{Bertrand:2004fq}. 
{\bf Figure~\ref{fig2}} shows that only ND-PAH, non complexed with siRNA, presents some toxicity, but at a high concentration with a IC50 of 0.1~$\mu$g$\,\mu$l$^{-1}$.  This effect might be due to cationic charges at the ND surface, as it has already been observed with polycation-coated silver nanoparticles~\cite{ElBadawy:2011}. When the cationic charge is neutralized by siRNA adsorption, the toxicity decreases. The relative decrease is larger in the case of PAH coating, since PAH is more toxic than PEI. 
However, for all ND-siRNA complexes the cell viability is larger than 70\%, even at the maximal concentration used. These results have to be compared to a 45\% toxicity which is observed after cell incubation with Lipofectamine siRNA lipoplexes at a dose corresponding to 50~nM siRNA final concentration (see Supporting Information) in conditions recommended by the supplier (serum free medium for 3~hours followed by 21 hours of culture in complete medium).

Cytotoxicity studies were also performed on cells expressing the EWS-Fli1 oncogene, i.e. the murine fibroblast NIH/3T3 EF cells and the  A673 human cells (see Supp. Info.). This later study allowed us to select the concentration range of siRNA-cationic NDs with a toxicity low enough not to interfere with the inhibition effect sought. Similar results were obtained,  i.e. concentration lower than 0.1~$\mu$g$\,\mu$l$^{-1}$.

\subsection{siRNA delivery into cells by cationic fluorescent nanodiamonds}

\begin{figure}[ht!]
\begin{center}
\includegraphics[width=0.8\textwidth]{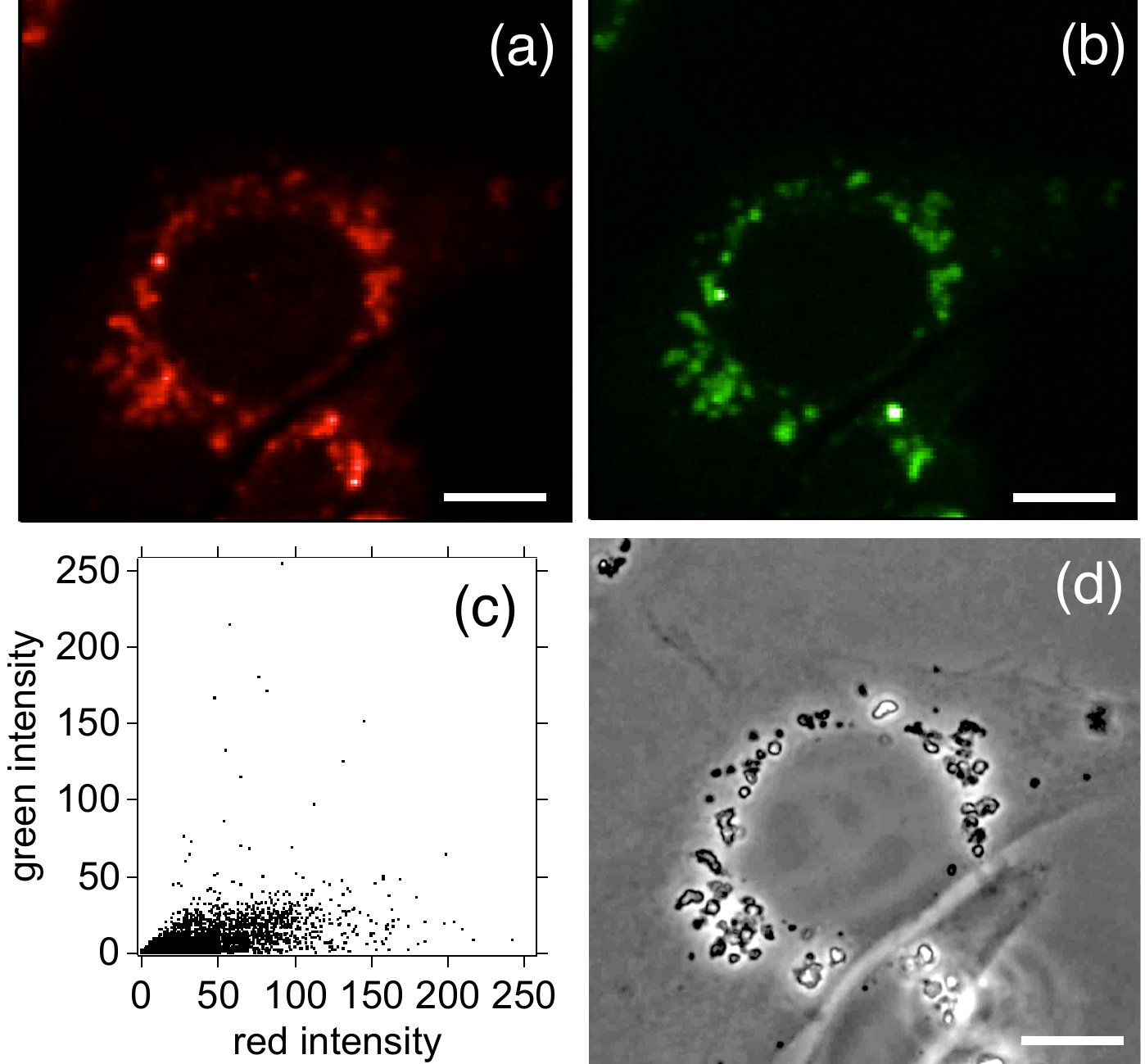}
\end{center}
\caption{\doublespacing Colocalization study between fND-PEI vectors (red channel,  {\bf (a)}), and siRNA labeled by FITC (green channel, {\bf (b)}), at siRNA:ND-PEI mass 1:75 ratio, in a NIH/3T3 EF cell. {\bf (c)} Scattered plot of the green vs. red intensity of images (a) and (b) after digitalization on 8-bits. {\bf (d)} phase contrast image of a cell showing that nanodiamonds are observed as aggregates in the perinuclear region. Observations made on cells fixed at $t=24$~h after the 4~h incubation with 50~nM siRNA, using a home-made confocal microscope~\cite{Faklaris:2009ec}, equipped with a $\times 60$, NA=1.4, oil microscope objective. Scale bar: 5~$\mu$m.}
\label{fig3}
\end{figure}

Cellular uptake efficiency and intracellular distribution of the siRNA-ND complexes have been evaluated using the fluorescence of FITC-labeled siRNA and that of fNDs. {\bf Figure~\ref{fig3}} displays a cell observed by confocal microscopy after 4~hour incubation with siRNA complexed to ND-PAH or ND-PEI. The signal in the red channel appears to be perfectly stable in time, which is a characteristic of fNDs fluorescence. As observed in previous studies nanodiamonds are localized in the perinuclear region and do not enter into the cell nucleus~\cite{Faklaris:2009ec}. 

For the given siRNA concentration of 50~nM, the ND-polycation mass was optimized to yield a sufficient amount of internalized nanoparticles as estimated from the intensity of the fND and FITC-labelled siRNA fluorescence signal (see Supp. Info.). The optimal siRNA:ND mass ratios were found to be 1:25 for ND-PAH and 1:75 for ND-PEI, corresponding to a $0.016~\mu$g$\,\mu$l$^{-1}$ and $0.048~\mu$g$\,\mu$l$^{-1}$ respectively. These ratios correspond to ND mass below the ones of the measured saturation values of 1:70 and 1:140 for ND-PAH and ND-PEI respectively, which means that the siRNA-ND complex is saturated in both cases. Moreover, no FITC-labelled siRNA is observed inside the cell if it is not adsorbed onto a cationic nanodiamond carrier (Figure~2(a) of Supp. Info.), as well known.
  
{\bf Figure~\ref{fig3}} shows a strong colocalization between FITC labelled siRNA and the fNDs, at time $t=24$~h after the 4~h incubation with siRNA:ND-PEI complex and change of medium to remove extra ND complexes. ImageJ \emph{JACoP} pluggin~\cite{Bolte:2006ws} was used to plot the green channel intensity (FITC) vs. the red channel one (fNDs). Despite difference in mean intensities of the two channels, {\bf Figure~\ref{fig3}(c)} scattered plot distribution indicates a some colocalization.  \emph{JACoP} pluggin also allows to quantify the fraction of red (resp. green) channel overlapping the green (resp. red) one, with Manders' coefficients $M_1$ (resp. $M_2$), taking as the threshold the onset of cellular autofluorescence on each channel ($3\sigma$ below the mean value of the autofluorescence intensity). Values $M_1=0.63$ and $M_2=0.96$ were obtained, confirming a high degree of colocalization. This results indicate that siRNA is not released in the form of large aggregates, but rather as small ones not detectable due to a fluorescence intensity of the same order of magnitude than the cell 
autofluorescence.

\begin{figure}[ht!]
\begin{center}
\includegraphics[width=\textwidth]{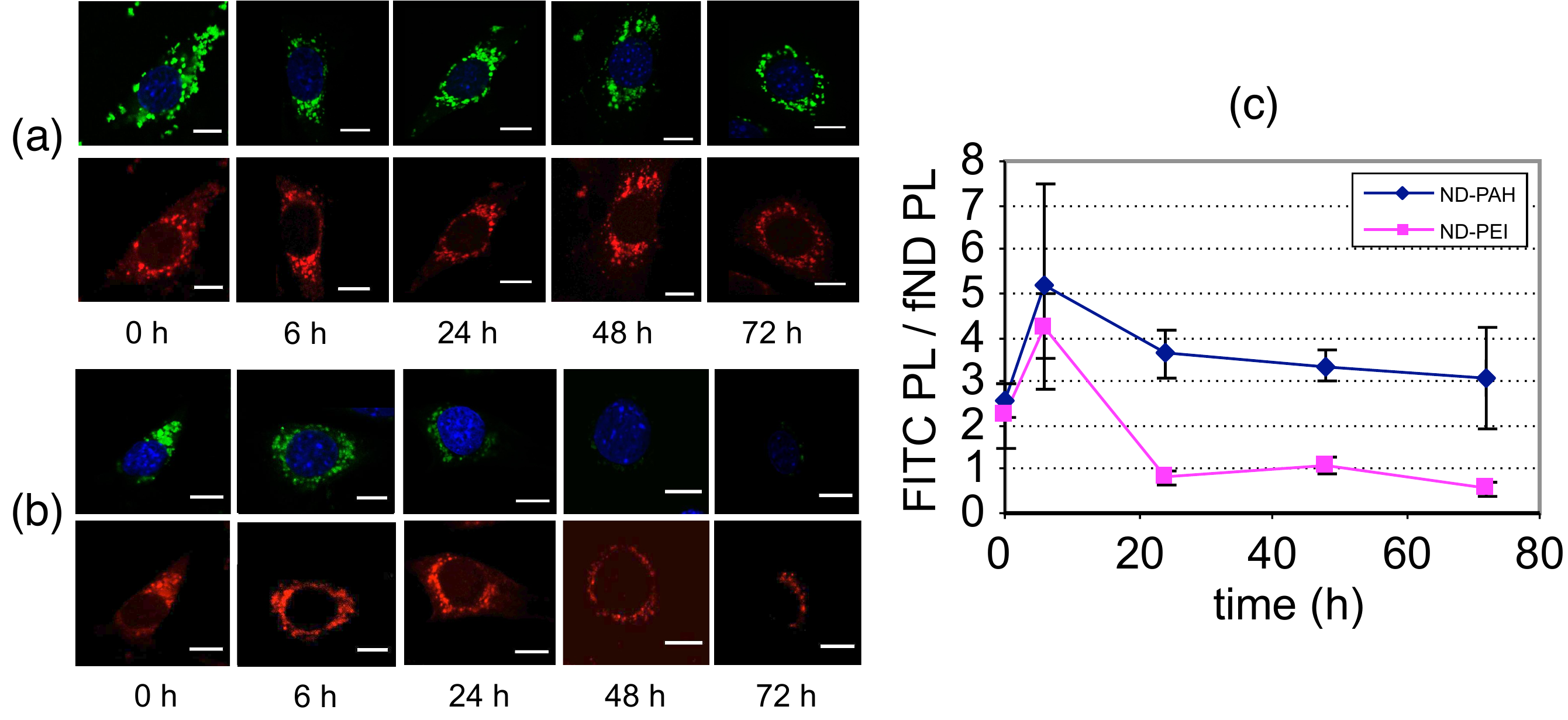}
\end{center}\caption{\doublespacing Kinetic study of the release of siRNA in NIH/3T3 EF cell after their vectorization by fND. siRNA is fluorescently labeled by FITC. Observations are done by confocal microscopy (Leica TSC SPE) at different times (0, 6, 24, 48, 72) hours after an initial  4~h incubation of the cells with 50~nM siRNA vectorized by either \textbf{(a)} ND-PAH or  \textbf{(b)} ND-PEI at (1:25; $1:75$) mass ratios. 
For each serie, the high panel (green channel) display the fluorescence of FITC (siRNA) and the down ones (red channel) the fluorescence of fNDs. Nucleus is labelled with DAPI (blue channel superimposed to the green one)). \textbf{(c)} Right figure: quantitative estimate of the siRNA release time, using the photoluminescence intensity of FITC (FITC PL) over the whole cell, normalized to that of fNDs (fND PL). The mean value of 5 to 20 cells is plotted. Scale bar: 10~$\mu$m.}
\label{fig4}
\end{figure}

In order to evaluate the dynamics of the release in cell of siRNA from the fNDs, we did confocal scans of cell cultures fixed at five different times (from 6 to 72~h), after the 4~h initial incubation duration with the NDs. For this study we used FITC-labeled siRNA complexed to fND-PAH (resp. fND-PEI) at the optimal mass ratio of 1:25 (resp. 1:75). At $t=6$~hours, {\bf Figures~\ref{fig4}(a)} and (b) show a strong signal of FITC at the same position than fND, indicating that siRNA are still bound to NDs. The FITC signal centered on the fNDs signal then drastically decreases in the case of PEI to stabilize at its lowest value at 24~h, while it only very slowly decreases for ND-PAH ({\bf Figure~\ref{fig4}(c)}).

Further studies have shown that NDs-polycation siRNA are internalized in Ewing sarcoma cells by endocytosis (data not shown). One hypothesis to account for the efficiency of polycations bearing amino groups as oligonucleotide delivery vehicles, is that they trigger the endosome disruption before the complex is transferred to the acidic lysosome compartment. The proposed mechanism relies on the protonation of the amino-polycation inside the cell due to its large pKa (pKa$\simeq 9$ for PEI~\cite{Choosakoonkriang:2003fg}), leading to an influx of counter-ions resulting in osmotic swelling followed by endosome membrane disruption and cytoplasmic release of the oligonucleotide-polycation complex~\cite{Creusat:2010ja,Xia:2009kb}.
Furthermore, in our case, the affinity of siRNA for ND-PAH being stronger than for ND-PEI, the desorption of siRNA from the nanodiamond carrier in the cytosol happens to be slower in the case of PAH coated ND.
 
\subsection{Inhibition of EWS-Fli1 by cationic ND vectorized siRNA}

\begin{figure}[ht!]
\begin{center}
\includegraphics[width=0.52\textwidth]{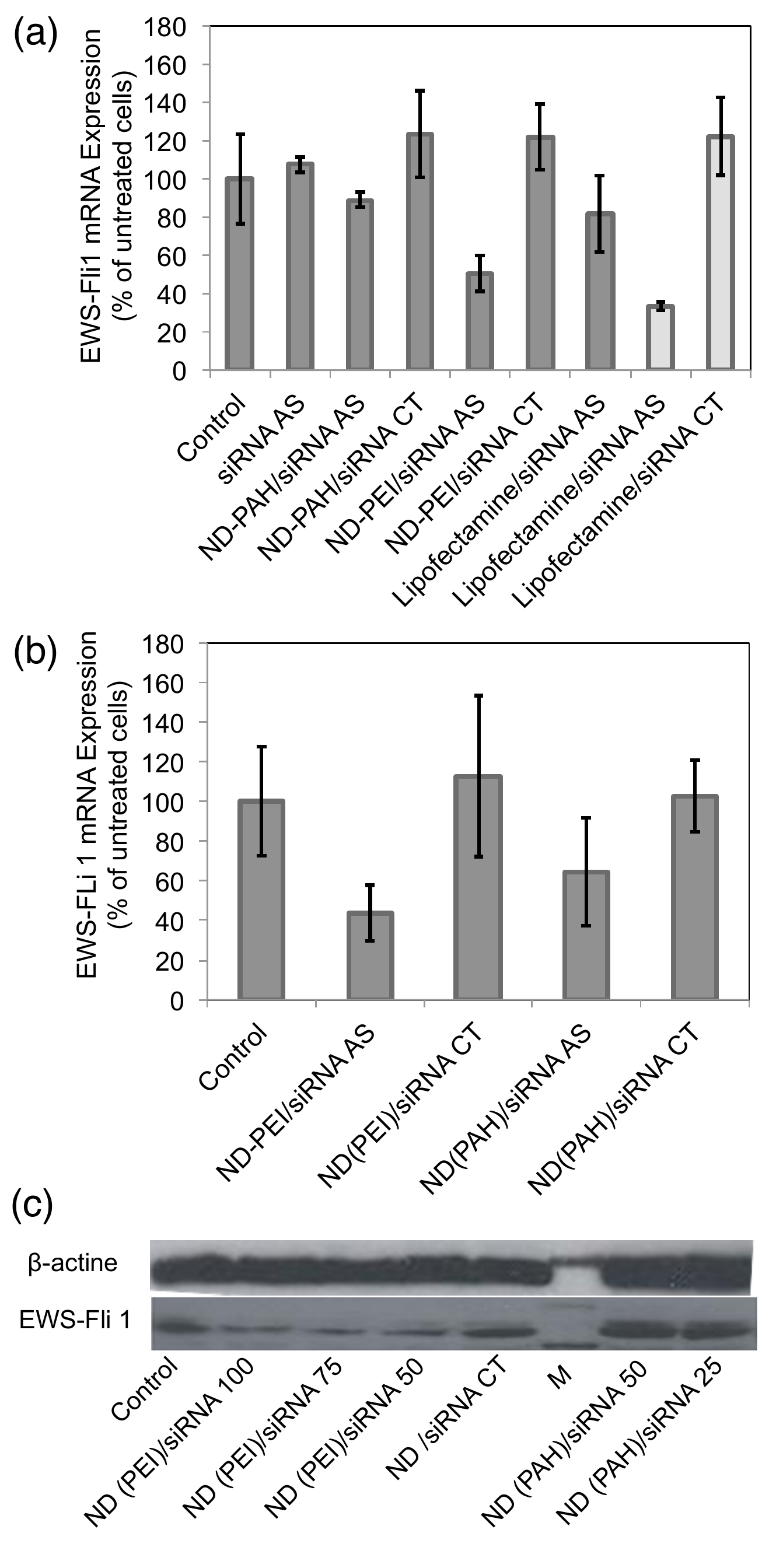}
\end{center}
\kern -6mm
\caption{RT-qPCR of EWS-Fli1 gene expression in \textbf{(a)} NIH/3T3 EF cells or in \textbf{(b)} A673 cells treated by 50~nM siRNA vectorized by ND-PAH at mass ratio 1:25, or by ND-PEI at ratio 1:75 or by lipofectamine as positive control. The EWS-Fli1 mRNA expression (controlled by the siRNA with the Anti-Sense action, siRNA AS) is normalized to the 18S RNA of the same sample used as a control (siRNA CT). Dark grey bars for cells treated in DMEM medium containing 10\% foetal bovine serum. Light grey bars in {\bf(a)} correspond to cells treated in OptiMEM medium \emph{without serum}. \textbf{(c)} Western blot assay of EWS-Fli1 expression in A673 cells incubated during 24~h with 50~nM siRNA:ND-PAH or ND-PEI at different weight ratios (1:$w$), where $w$ value is indicated at the end of the legend. ``M'' stands for medium alone, and the ``Control'' experiment (first data from the left) was done on untreated A673 cells.}
\label{fig5}
\end{figure}

The efficiency of siRNA vectorized by cationic NDs has been evaluated by their ability to interfere destructively with the targeted EWS-Fli1 mRNA involved in Ewing sarcoma. After cell treatment with 50~nM EWS-Fli1 siRNA vectorized either by ND-PAH or ND-PEI, we have measured by qPCR the EWS-Fli1 mRNA ratio in treated and untreated cells as described in the Experimental Section. The capacity of siRNA to inhibit mRNA expression was tested with the two cell lines expressing EWS-Fli1: NIH/3T3 EF and A673. {\bf Figure~\ref{fig5}(a)} shows that 50~nM free siRNA do not inhibit EWS-Fli1 expression. 
When it is vectorized by either ND-PAH or ND-PEI, EWS-Fli1 mRNA expression is inhibited, with the maximum efficiency of 50\% for ND-PEI in the case of NIH/3T3 EF cells and resp. 55\% with A673 cells ({\bf Figure~\ref{fig5}(b)}). ND-PAH appears to be a less efficient transfection agent. In the same conditions a control siRNA, with no anti-sens action, gives no inhibition effect. 

When Lipofectamine$\textsuperscript{\textregistered}$ is used to vectorize the same quantity of siRNA in a free serum medium, we obtained 65\% inhibition of the targeted mRNA, but associated to a 50\% cytotoxic effect compared to only 20\% for ND-PAH or ND-PEI. Moreover, if the cells are treated by Lipofectamine in serum containing medium, the inhibition of EWS-Fli1 mRNA expression is only 20\% to be compared with 50\% for ND-PEI vectorized siRNA, proving that ND-PEI is a better transfection agent for siRNA in physiological conditions.
 
To confirm the specific EWS-Fli1 inhibition by nanodiamond-vectorized siRNA, we have determined the protein expression by a Western blot assay, using the A673 cell line. As shown on {\bf Figure~\ref{fig5}(c)} a specific inhibition of EWS-Fli1 oncogenic protein takes place with siRNA vectorized by ND-PEI  at a mass ratio of 1:75. No effect is observed on $\beta$-actin used as house-keeping protein. By using siRNA:ND-PAH at a mass ratio of 1:25 inhibition of EWS-Fli1 protein expression can hardly be observed, which is consistent with RT-qPCR data.

The larger inhibition efficiency of siRNA:ND-PEI compared to siRNA:ND-PAH is also in perfect agreement with a slower release of siRNA in the later case, as observed in confocal microscopy ({\bf Figure~\ref{fig4}}).

\section{Conclusions}
We have synthesized a fluorescent nanodiamond vector with two different cationic coatings, for siRNA delivery into Ewing sarcoma cell in culture. 
This vector shows advantages compared to the usual transfection agent lipofectamine, inducing a larger efficiency in inhibiting the EWS-Fli1 expression, combined with a lower toxicity to the cell in serum supplemented medium. 
We observed that the larger adsorption affinity of siRNA onto PAH-coated NDs than onto PEI-coated NDs results in a much slower dissociation of the siRNA:ND-PAH complex than of the siRNA:ND-PEI ones, and hence a lower siRNA-associated biological activity. Moreover, we showed that the ND-PEI carrier is less toxic than the ND-PAH ones.

This study shows that the efficient delivery of an oligonucleotide by a cationic solid nanoparticle consists in a compromise between (i) a sufficiently strong adsorption of the biomolecule onto the particle to go through the cell membrane without loss of material, (ii) the dissociation of the complex on the timescale of a cell division cycle, and (iii) a low cellular toxicity. Furthermore a careful selection of the cationic polymer may even serve to control the release kinetics of the siRNA.

The next step will be to test the activity of the siRNA delivered by the cationic NDs \emph{in vivo} and study the elimination of the vector. The biodistribution, the toxicity, and pharmacokinetics studies of this new vector will be facilitated by its fluorescence properties.

\section{Experimental Section}\label{sect:exp}

\emph{ND-polycation complex formation and characterization (ND surface functionalization by polycation, siRNA adsorption):}
This procedure was optimized using as-received, non fluorescent nanodiamonds first, of an average size of 50~nm (SYP 0-0.05, 50~nm; Van Moppes, Geneva, Swizerland). These particles are cleaned in a strongly oxidative acid mixture by the manufacturer. Their dispersion in water yields a colloidal suspension with negative zeta potential values as reported in {\bf Table~\ref{table:01}}. This is attributed to carboxylic acid groups at the diamond surface~\cite{Xu:2005ws}, further confirmed by FTIR analysis displaying a broad band centered at 1775~cm$^{-1}$ (C=O stretching bond)~\cite{Chung:2006ib}.

As-received NDs were dispersed in deionized water to achieve a 1~g\,l$^{-1}$ concentration. The suspension was sonicated 3~h, at 300~W (Vibra-Cell with a 3~mm stepped microtip) to ensure  de-agglomeration.
We used branched low molecular weight (800~Da) Polyethylenimine (PEI), ethylenediamine end-capped (Sigma-Aldrich Ref.~408719) and Poly(allylamine hydrochlolide) (PAH, Sigma-Aldrich Ref. 283215) polycations. To adsorb the polymers on NDs, 1~g\,l$^{-1}$ NDs water solution was added dropwise to a PAH solution (v/v) (1~g\,l$^{-1}$, 3~mM NaCl) or to PEI solution (at the ND particle:PEI molar ratio 1:10, like in \citet{Zhang:2009hc}) and then strongly sonicated at power 300~W, for 15 min. The mixtures were stirred overnight at room temperature, then washed 3 times with deionised water and centrifugated at 25,000~rpm for 1~h (\emph{Optima XL90} Ultracentrifuge, \emph{50Ti} rotor, Beckman Coulter, Inc., USA). The ND-Polymer pellet was redispersed in water, and the concentration was determined by weighting after lyophilization of 1~ml of the solution.

The size of the ND and ND-polycation complex were determined by dynamic
quasi-elastic light scattering (\emph{Nano ZS}, Malvern Instrument, UK). The measurements were done at 20$^\circ$C, using an aqueous solution of NDs at the concentration of $20~\mu$g\,ml$^{-1}$. The zeta potential was determined with the same instrument using a solution of NDs in 1~mM NaCl at the concentration of $80~\mu$g\,ml$^{-1}$. Both the size and the zeta potential values are reported in {\bf Table~\ref{table:01}} and correspond to the average of three measurements.

After coating by PAH or PEI, the quantity of amine associated to the NDs was determined by the Kaiser method~\cite{KAISER:1970bv} as described by Vial and al.~\cite{Vial:2008jw}. Briefly, we added to 100~$\mu$l of dried amino-substituted NDs, 75~$\mu$l of phenol/ethanol solution (8~g in 2~ml), 100~$\mu$l pyridine/KCN solution (98~ml pyridine in 2~ml of 1~mM KCN) and 75~$\mu$l ninhydrin/ethanol solution (1~g in 20~ml). The solution was heated at 100$^\circ$C for 5~min and diluted with the addition of 2~ml of ethanol. The solution was centrifugated during 5 min at 13,000~rpm to remove the NDs particles. The amine concentration in the supernatant is inferred from the light absorption coefficient at the wavelength of 570~nm. The quantity of amine present is expressed in $\mu$mole\,g$^{-1}$ of ND suspension using $\varepsilon$=19,700 M$^{-1}$cm$^{-1}$ as the molar extinction coefficient.

PEI or PAH coated nanodiamonds at increasing concentration (0 to 0.4~$\mu$g$\,\mu$l$^{-1}$) were incubated with a constant concentration of siRNA (5~ng\,$\mu$l$^{-1}$ for 15~min at room temperature. In order to determine the remaining free siRNA
concentration, the solution was first centrifugated at 13,000~rpm
during 30~min, then 10~$\mu$l aliquot of supernatant was mixed with 10$\mu$l ethidium bromide (EtBr) at 2~$\mu$g$\,\mu$l$^{-1}$ concentration. The EtBr fluorescence was quantified under UV illumination with a gel imaging and analysis system (\emph{InGenius}, Syngene, UK) using 5$\mu$l drop of the mixture.

\emph{Cellular studies (cell culture and viability, cell transfection):}
The NIH/3T3 cells expressing human EWS-Fli1 oncogene is a generous gift from Dr. J. Ghysdael (Institut Curie, Orsay, France). These cells were grown in DMEM medium (\emph{Gibco}$\textsuperscript{\textregistered}$, Invitrogen Corp., USA) containing 10\% newborn calf serum (Gibco), 1\% penicillin-streptomycin antibiotics (Gibco) and 2.5~$\mu$g$\,\mu$l$^{-1}$ Puromycin (Sigma-Aldrich, USA). Incubation was performed at 37$^\circ$C, 5\% CO$_2$ in a moist atmosphere. A673 human Ewing sarcoma cells were a generous gift from Dr. Elizabeth R. Lawlor (University of Michigan, USA) and were grown in DMEM medium supplemented by 10\% of foetal bovine serum (Gibco) and 1\% penicillin-streptomycin antibiotics (Gibco). 

Cell viability in the presence of polycation-coated NDs was determined on regular NIH/3T3 cell line by MTT test (Sigma-Aldrich). Cells, plated within 96 wells (\emph{TPP}, Dominique DUTSCHER SAS, France) with $2\times10^4$ cells in 100~$\mu$l of full medium per well, were incubated with nanodiamonds at indicated doses during 24~h at 37$^\circ$C. Then, 10~$\mu$l of 5~mg\,ml$^{-1}$ MTT in PBS was added to the culture during 2~h. Cells were then lysed by 100~$\mu$l of 10\% sodium docecyl sulfate and 10~mM HCl solution over night. The absorbance of the  formazan produced was measured at 570~nm wavelength on a plate reader (\emph{MRX2}{\footnotesize\texttrademark}, Dynex Technologies, USA). Results are expressed in percentage using untreated cells as the residual absorbance reference.

The effect of siRNA was measured after its cell transfection with ND-PAH or ND-PEI. 
The siRNA sequences were designed to target the oncogene junction EWS-Fli1 in the chimeric mRNA (nucleotides 822-842). 
The siRNAs with the following sequences were purchased from Eurogentec (Li\`eges, Belgium):
\\ siRNA antisense: sense strand 
5'-r(GCUACGGGCAGCAGAACCC)d(TT)-3', \\
antisense strand 5'-r(GGGUUCUGCUGCCCGUAGC)d(TG)-3' ;\\ 
siRNA control: sense strand 5'-r(GCCAGUGUCACCGUCAAGG)d(AdG)-3', \\ 
and antisense strand 5'-r(CCUUGACGGUGACACUGGC)d(TdT)-3'.\\
The sense and antisense strands were hybridized at 20~$\mu$M concentration in annealing buffer (Eurogentec) by heating 5~min at 95$^\circ$C followed by 1~h at 37$^\circ$C.
NIH/3T3 EF or A673 cells were seeded into 12~wells plates at the density of $1.5\times10^5$ cells per well in 500~$\mu$l of corresponding medium 24~h before transfection. Different mass ratios of ND/siRNA antisens and control were prepared as following. siRNA  were prepared at 250~nM concentration in 100~mM NaCl, 10~mM Hepes buffer (pH=7.3). A volume of 40~$\mu$l of this solution was mixed with 60~$\mu$l of ND-PAH or ND-PEI prepared in 100~mM NaCl, 10~mM Hepes buffer, at different concentrations of nanodiamonds. The mixture was incubated during 15~min at room temperature to form the ND/siRNA complexes. The medium in each well was finally replaced with 400~$\mu$l of fresh culture medium containing serum and 100~$\mu$l of the complexes. Cells where then incubated for the time indicated in each experiment.

\emph{Fluorescence Microscopy (synthesis of fNDs, cell imaging):}
fNDs were prepared according to Dantelle et al.~\cite{Dantelle:2010hc} procedure. Briefly, vacancies were created in diamond nanocrystals (size$<50$~nm, SYP0-0.05, Van Moppes) by a electron beam irradiation (energy: 13~MeV). NV centers were formed upon annealing in vacuum (800$^\circ$C, 2~hours), leading to the migration of the vacancies in the diamond matrix and their stabilization in adjacent sites of a nitrogen impurity in substitution. The graphitic shell formed on ND surface during this process was finally removed by air oxidation~\cite{Rondin:2010dn}. The NV color center (hence fNDs) emission spectrum spans from a wavelength of 600 to 750~nm.

The cells were plated in 12-wells plates ($1.5\times10^5$ cells and 500~$\mu$l of medium per well) with a cover glass (18~mm in diameter) at the bottom of each well, and cultured during 24~h. fND-PEI or fND-PAH were complexed with fluorescently labelled siRNA (3'-FITC-siRNA antisense, from Eurogentec) at the concentration of 50~nM siRNA, and at different siRNA:ND weight ratios from 1:25 to 1:75. The cell medium was replaced by 100~$\mu$l of the fND-siRNA complex solution, completed by 400~$\mu$l of serum free OptiMEM medium (Gibco) in each well. After 3 h of incubation, cells were washed by PBS, and then fixed with paraformaldehyde 4\% in PBS during 20 min at room temperature. After two washes with PBS, slides were mounted with Dapi Fluoromount G (Southern Biotech) and then observed with confocal microscopy (TCS SPE Leica coupled to a DMI 4000B Leica microscope). The green channel corresponds to the 501-543~nm spectral region, while the red one is associated to the 654-702~nm wavelength range. For the quantification of the kinetics of siRNA release, the intensities of both channels was calculated using NIH ImageJ software. It can be assumed that  that only FITC-labeled siRNA bound to the fNDs contributes to the green intensity, since when siRNA were released they spread all over the cytosol, and FITC signal at each pixel could not overcome autofluorescence
intensity.
Colocalization studies of fNDs and siRNA were realized on a home-made scanning stage confocal microscope described in Faklaris et al.~\cite{Faklaris:2009ec} equipped with single-photon counting detectors.

\emph{Biological activity tests(Real-Time quantitative PCR, Inhibition of Proteins synthesis) :}
The total RNA extraction was performed as following. After 24~h incubation, cells were washed with PBS and lysed with 800~$\mu$l TRIzol (Invitrogen). Note that this procedure only keep live cells, dead cells being washed away. Then, 160~$\mu$l of chloroform was added, and the mixture was centrifugated at 13,000~rpm for 15~minutes. A volume of 300-350~$\mu$l of the aqueous phase was mixed with the same volume of isopropanol, incubated for 15 min at room temperature and centrifuged at 13,000 rpm for 15 min at 4$^\circ$C. The pellet was washed twice with 70\% ethanol, dried at room temperature, and then dissolved in 10~$\mu$l of water containing 0.5~unit\,$\mu$l$^{-1}$ RNasin$\textsuperscript{\textregistered}$ (Promega Corp., USA) ribonuclease inhibitor. The total amount of RNA was quantified using spectrophotometry (\emph{UV1605}, Shimadzu Corp., Japan) at the wavelength of 260~nm.
Reverse transcription was performed on 1.5~$\mu$g of total RNA by adding 2~$\mu$l of Random Hexamers at 50~$\mu$g$\,\mu$l$^{-1}$ (Promega), and heating at 65$^\circ$C during 5~min. Then RNA was incubated with 0.5~$\mu$l M-MLV Reverse Transcriptase 200~unit/$\mu$l, 0.5~$\mu$l dNTP 20~mM, 0.5~$\mu$l RNasin 40~unit/$\mu$l and 4~$\mu$l of M-MLV RT 5$\times$ reaction buffer (Promega) for 1~h at 42$^\circ$C.
Quantitative PCR  was performed using SYBR GreenER qPCR SuperMix (Invitrogen). EWS-Fli1 gene was amplified using EWS-Forward Primer: 5'-AGC AGT TAC TCT CAG CAG AAC ACC-3' and Fli1-reverse primer: 5'-CCA GGA TCT GAT ACG GAT CTG GCT G-3' (Eurogentec). One $\mu$l of each primer at 10 ~$\mu$M was mixed to 5~$\mu$l of cDNA diluted to 1/20 (v/v) in 25~$\mu$l final volume. Samples were amplified with 45 cycles using 7900HT Fast Real-Time PCR System (Applied Biosystems, USA) as follows: 2~min incubation at 50$^\circ$C, 10 min at 95$^\circ$C, followed by 45 cycles at 95$^\circ$C during 15 s and 60$^\circ$C during 1~min. Human 18S gene was used as a PCR reference gene, using 18S forward primer 5'-CGT TCA GCC ACC CGA GAT-3', and 18S reverse primer 5' TAA TGA TCC TTC CGC AGG TT-3'. The quantification cycle ($C_{\rm q}$) is between 10-16 for 18S gene and 20-24 for EWS-Fli1 gene. The results are expressed as a percentage compared to untreated cells. The whole process starts with live cells, and is therefore insensitive to the toxicity of the siRNA:ND-polycation complex, although a low toxicity is desirable for a better quantification.

Proteins extracts were obtained from $3\times10^5$ cells per well, grown in 6 wells plates and incubated during 24~h with different weight ratio of ND/siRNA. The cells are lysed using 200~$\mu$l RIPA Buffer (50~mM Tris pH=7.4, 150~mM NaCl, 1~mM EDTA, 10\% Glycerol, 0.5\% NP40 and complete protease inhibitor (Roche, Germany). After cell lysis, extracts were cleared by centrifugation at 13,000 rpm during 15~min at 4$^\circ$C.
Proteins concentration was determined using BCA protein assay (Thermo Scientific Pierce, USA). Proteins were separated by NuPAGE 10\% Bis-Tris Gel (Invitrogen) and transferred to BA-S 85 nitrocellulose membrane (Schleicher \& Schuell, GE Healthcare, UK). After blockage phase with 5\% milk in PBS containing 0.1\% Tween-20, the membrane was probed over night at 4$^\circ$C with rabbit anti-Fli1 (c term) antibody (Santa Cruz Biotechnology, USA) diluted to 1/500 in 2.5\% of milk in PBS containing 0.1\% Tween-20. The first antibody was detected using anti-rabbit antibody bounded to peroxidase (GE Healthcare) diluted to 1/1500 and revealed with chemiluminescence Kit (Amersham{\footnotesize\texttrademark}~ECL{\footnotesize\texttrademark}, GE Healthcare). Actin protein was used as housekeeping gene and was detected using anti actin antibody (Sigma). 

\section*{Acknowledgements}
We thank Orestis~Faklaris, Huan-Cheng Chang, Catherine Durieu and Eric Le~Cam for fruitful discussions. We are grateful to F.~Lain\'e, F.~Carrel and P.~Bergonzo for providing us the access to the SAPHIR electron facilities from the CEA-LIST (CEA Saclay, Gif-sur-Yvette, France), and for their help in the preparation of fluorescent nanodiamonds. We thank Olivier Duc for assistance in confocal imaging at Institut Gustave Roussy, and Unit\'e de G\'enomique Fonctionnelle facilities at Institut Gustave Roussy for his help in Q-PCR analysis.
This work has been supported by the Region Ile-de-France in the framework of C'Nano IdF. C'Nano IdF is the nanoscience competence center of Paris Region, supported by CNRS, CEA, Ministry of Higher Education and Research and Region Ile-de-France, France. 


\bibliographystyle{angew}
\bibliography{ref_rev}

\bigskip
\singlespacing
\rightline{Received: xxx}
\rightline{Revised: xxx}
\rightline{Published online on xxx}

\newpage

\section*{Tables and Figures}
\begin{table}[h!]
  \caption{\doublespacing Physico-chemical characteristics of the two types of ND-polycation complex in aqueous suspension compared to untreated ND.}
  \label{table:01}
  \medskip
  \begin{tabular}{ccccc}
    \hline
Type of nanodiamonds & Size & Zeta Potential & pH & Amine content\\
& (nm) & (mV) & & ($\mu$moles/g)\\
\hline
ND	 	&  54		&-28 	&-- 	&--\\
ND-PAH	&133	&+31 	&5.6	& 274\\
ND-PEI	&123	&+27 	&6.7 & 173\\
    \hline
  \end{tabular}
\end{table}

\vskip 2cm

\clearpage

\makeatletter \renewcommand{\thefigure}{S\@arabic\c@figure} 
\setcounter{figure}{0}

\section*{Supporting information}

\subsection*{Viabillity of cells expressing EWS-Fli1 oncogene after incubation with siRNA:ND.}
In addition to cell viability study on normal NIH/3T3 cells proving that the siRNA:ND-polycation complex only weakly impacts the cell viability (see main text), we also studied the viability of cells expressing EWS-Fli1 oncogene to proved that the viability is sufficiently high not to impair the measurement of the efficiency of siRNA inhibition. 
 
\begin{figure}[ht!]
\centerline{\includegraphics[width=0.7\textwidth]{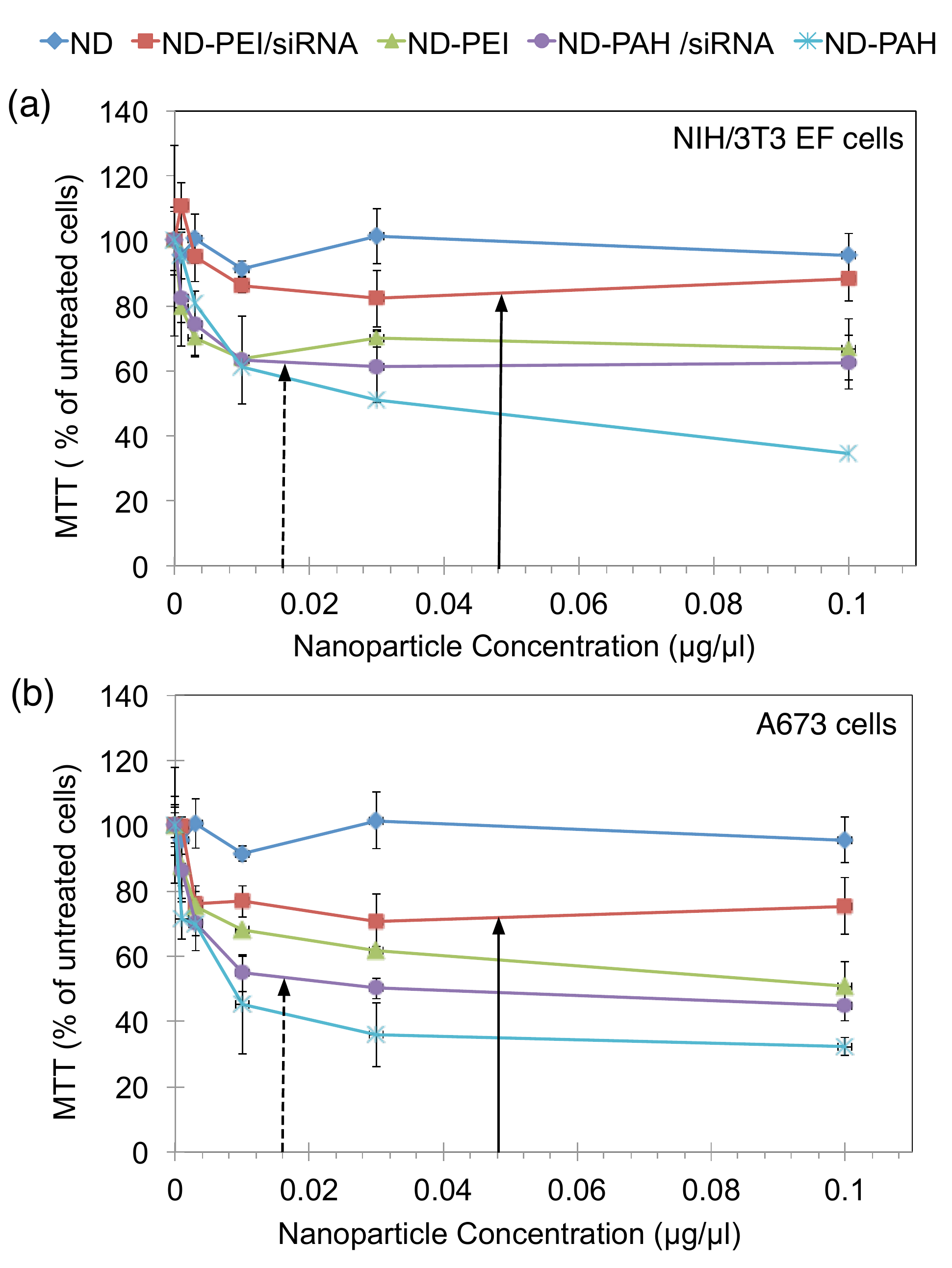}}
\caption{Cell viability measurements by MTT assay. {\bf (a)} NIH/3T3 EF cells or {\bf (b)} A673 cells were incubated with ND and ND coated with cationic polymers with doses from 0.001 to $0.1~\mu$g$\,\mu$l$^{-1}$ for 24~h, and siRNA concentration of 50~nM. 
Let us first point out that the free antisens siRNA at concentration 50~nM does not impede the cellular viability. Viability results are presented in percent as the ratio of MTT signal for treated cells on the one of untreated cells, i.e. cells not incubated with ND or ND complexes. Vertical dashed and solid arrows indicate the concentrations of ND-PAH (0.016~$\mu$g$\,\mu$l$^{-1}$) and ND-PEI (0.048~$\mu$g$\,\mu$l$^{-1}$) respectively, which are used in EWS-Fli1 gene inhibition experiment. }
\label{fig:figure1Sup}
\end{figure}
Murine fibroblast stably expressing human EWS-Fli1 oncogene (NIH/3T3 EF) and human Ewing sarcoma cells (A673) were treated by increasing concentrations of NDs, cationic polymer coated NDs and siRNA bound to the cationic NDs. After 24 hours incubation, survival of NIH/3T3 EF cells was determined by a MTT assay. 
{\bf Figure~\ref{fig:figure1Sup}(a)} shows that ND alone present a low toxicity in the experimental conditions. Adsorption of cationic polymers on ND increases their toxicity with an IC50 of 30~$\mu$g$\,\mu$l$^{-1}$ for ND-PAH and a toxicity of about 30\% for ND-PEI at concentration higher than 15 $\mu$g$\,\mu$l$^{-1}$. Surprisingly, the toxicity does not increase with larger ND-PEI concentrations, which might be due to the aggregation of the nanoparticles. 
When siRNA are adsorbed to ND-PAH or ND-PEI (at 50~nM concentration), the toxicity decreases to around 20\%, for concentrations larger than 10~$\mu$g$\,\mu$l$^{-1}$.
Similar results have been obtained with human A673 Ewing cells ({\bf Figure~\ref{fig:figure1Sup}(b)}). ND alone displays a low toxicity (less than 10\%). Cationic ND-PAH or ND-PEI are more toxic with IC50s of 8~$\mu$g$\,\mu$l$^{-1}$ and 90~$\mu$g$\,\mu$l$^{-1}$ respectively. When siRNA are adsorbed onto the ND particles, we observe a decrease of their toxicity with an IC50 of 20~$\mu$g$\,\mu$l$^{-1}$ for ND-PAH and a constant 20\% toxicity for concentrations higher than 5~$\mu$g$\,\mu$l$^{-1}$ for ND-PEI.

\begin{figure}[ht!]
\centerline{\includegraphics[width=0.6\textwidth]{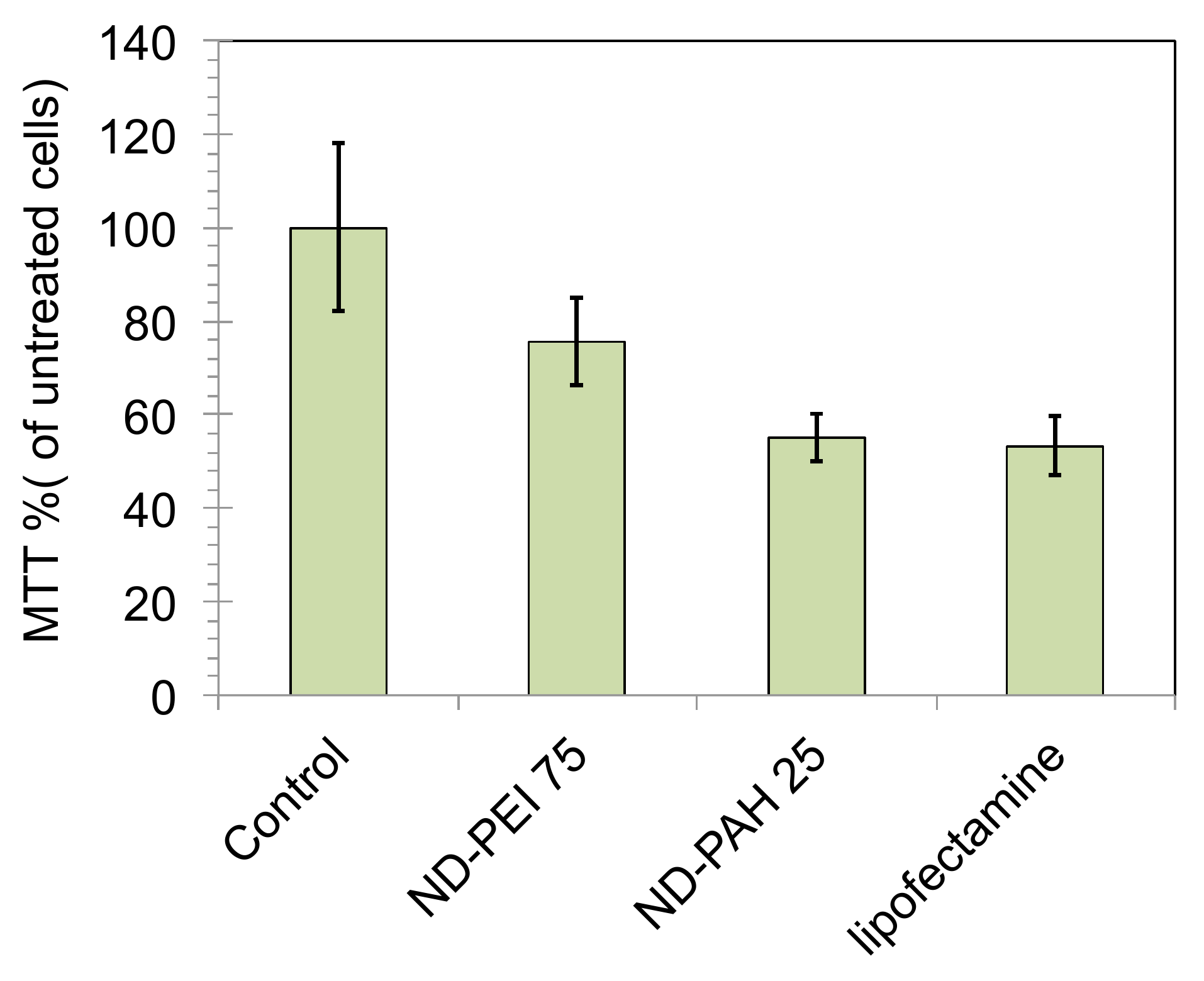}}
\caption{Comparison of NIH/3T3 EF cell viability measured by MTT assay between lipofectamine (no serum supplemented culture medium) and ND-polycation carriers, with siRNA at 50~nM concentration. siRNA to ND-polycation mass ratios are 1:75 and 1:25 for siRNA:ND-PEI and siRNA:ND-PAH respectively, corresponding to 0.016~$\mu$g/$\mu$ and 0.048~$\mu$g$\,\mu$l$^{-1}$ for the 50~nM siRNA concentration.}
\label{fig:figure2Sup}
\end{figure}

Note that the efficient inhibition of EWS-Fli1 oncogene was observed at ND-PEI concentration of 0.048~$\mu$g$\,\mu$l$^{-1}$ (solid vertical arrow in {\bf Figure~\ref{fig:figure1Sup})}, at which a viability larger than 70\% for both cell lines is measured. Similarly, a viability fraction larger than 55\% is obtained in the case of the ND-PAH carrier at the concentration of 0.016~$\mu$g$\,\mu$l$^{-1}$ used in the inhibition experiment (vertical dashed arrow). At these optimal concentration the lipofectamine carrier yields a cell viability of about 55\% for NIH/3T3 EF cells ({\bf Figure~\ref{fig:figure2Sup}}).

\subsection*{Optimal mass of ND-polycation for a sufficient amount of siRNA:ND-PAH/PEI internalized in NIH/3T3 EF cells in culture}
\begin{figure}[ht!]
\centerline{\includegraphics[width=\textwidth]{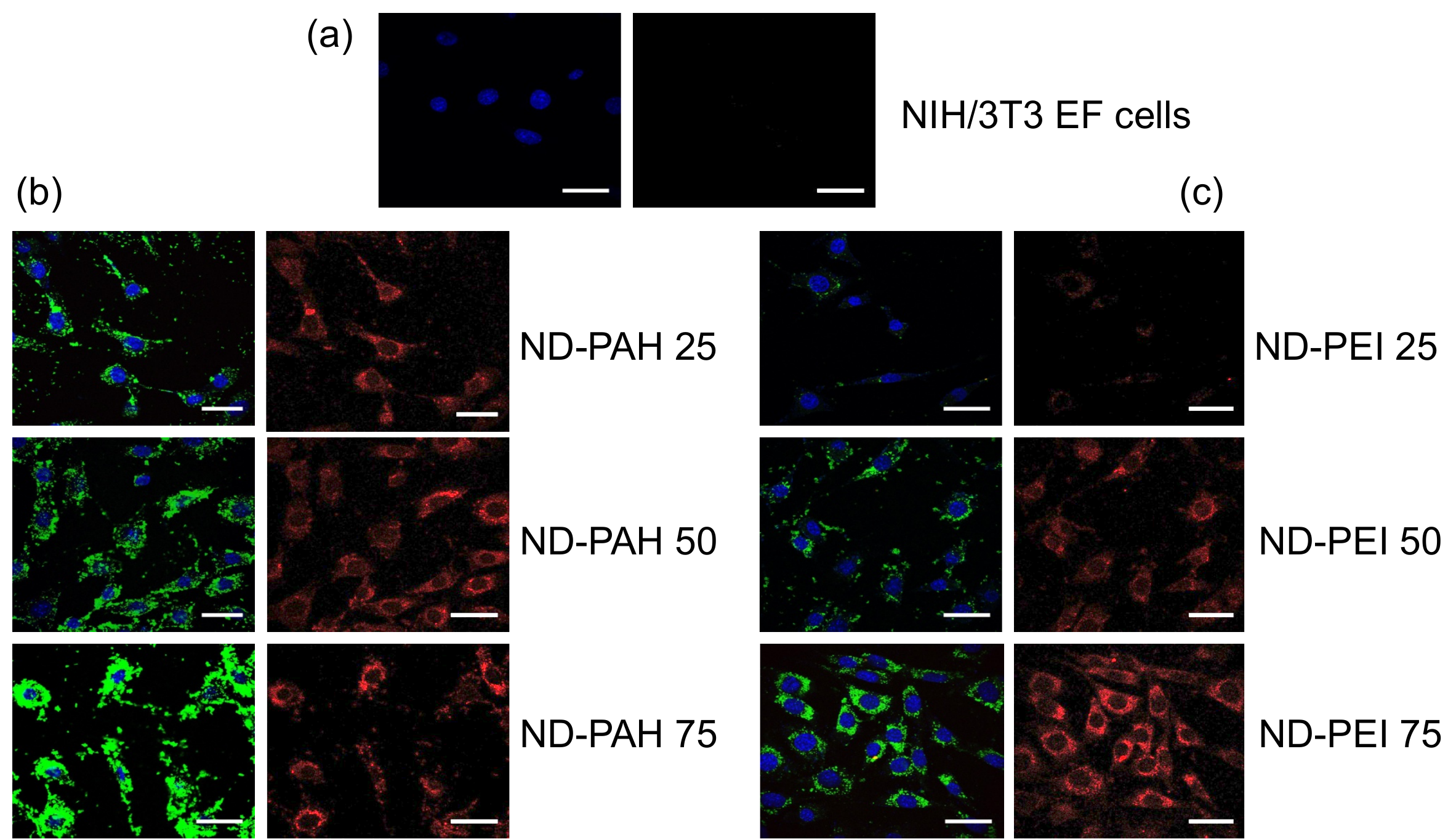}}
\caption{Analysis by confocal microscopy of cell distribution of FITC-labeled siRNA vectorized by fluorescent NDs in NIH/3T3 EF cells. The cells are incubated during 4 hours \textbf{(a)} with 50~nM siRNA alone, \textbf{(b)} with  50~nM siRNA vectorized by fND-PAH or \textbf{(c)} with 50~nM siRNA vectorized by fND-PEI, at various siRNA:ND mass ratio $1:w$, with $w=25, 50$ or 75 corresponding to 0.016, 0.032 and $0.048~\mu$g$\,\mu$l$^{-1}$ ND-polymer concentrations repectively. Observations are done on fixed cells. The cell nuclei are colored with DAPI. Each of the 6 experiments is visualized twice for the same field:  siRNA (left panel, green and blue channel) and fND (right panel, red channel).Scale bar: 20~$\mu$m.}
\label{fig:figureSup}
\end{figure}
We determined the optimal mass of ND-polycation for a sufficient amount of siRNA:ND-PAH/PEI internalized in NIH/3T3 EF cells in culture using confocal microscopy raster scans of fNDs and FITC-labelled siRNA fluorescence. As expected, {\bf Figure~\ref{fig:figureSup}(b)} and {\bf Figure~\ref{fig:figureSup}(c)} show that the intracellular content of siRNA increases with the fND carrier added mass, expressed as a ratio 1:$w$, for a given constant amount of siRNA (50~nM). Moreover, the absence of FITC fluorescence in {\bf Figure~\ref{fig:figureSup}(a)} proves that siRNA alone cannot enter the cells, justifying the need for a carrier.

From a qualitative observation of {\bf Figure~\ref{fig:figureSup}}, one concludes that 1:25 and 1:75 are the best siRNA:ND-PAH and siRNA:ND-PEI mass ratios respectively, among the one tested, to achieve a sufficient amount of internalized siRNA. The corresponding ND-PAH and ND-PEI concentrations of 0.016~$\mu$g$\,\mu$l$^{-1}$ and 0.048~$\mu$g$\,\mu$l$^{-1}$ respectively, were then selected to test EWS-Fli1 gene inhibition at the same siRNA 50~nM concentration.
\end{document}